\documentclass[conference, 10 pt]{IEEEtran}
\IEEEoverridecommandlockouts    

\usepackage{cite}

\ifCLASSINFOpdf
\usepackage[pdftex]{graphicx}
\else
\fi
\usepackage{bbold}
\usepackage{dsfont}
\usepackage[cmex10]{amsmath}
\usepackage{algorithmicx}
\usepackage{algpseudocode}
\usepackage{algorithm}
\usepackage{float}
\newfloat{algorithm}{t}{lop}

\algnewcommand\algorithmicinput{\textbf{INPUT:}}
\algnewcommand\INPUT{\item[\algorithmicinput]}

\algnewcommand\algorithmicoutput{\textbf{OUTPUT:}}
\algnewcommand\OUTPUT{\item[\algorithmicoutput]}

\algnewcommand\algorithmicinit{\textbf{Initialization:}}
\algnewcommand\Init{\item[\algorithmicinit]}

\newcommand{\Nc}{N}

\newcommand{\rr}{\boldsymbol{r}}
\newcommand{\ssv}{\boldsymbol{s}}

\newcommand{\mathcalB}{B}

\newcommand{\mathcalK}{S}
\newcommand{\mathcalM}{M}

\newtheorem{theorem}{Theorem}

\newtheorem{corollary}{Corollary}[theorem]

\begin{document}
%
\title{Traffic Driven Resource Allocation in Heterogenous Wireless Networks}

\author{\IEEEauthorblockN{Binnan Zhuang, Dongning Guo and Michael L. Honig\\}%
\thanks{This work was supported by a gift from Futurewei Technologies, Inc.\
and the National Science Foundation under Grant No.~CCF-1018578.}%
\IEEEauthorblockA{Department of Electrical Engineering and Computer Science\\
Northwestern University\\
2145 Sheridan Road, Evanston, IL 60208 USA}}


%


\maketitle

\begin{abstract}
Most work on wireless network resource allocation use physical layer performance such as sum rate and outage probability as the figure of merit. These metrics may not reflect the true user QoS in future heterogenous networks (HetNets) with many small cells, due to large traffic variations in overlapping cells with complicated interference conditions. This paper studies the spectrum allocation problem in HetNets using the average packet sojourn time as the performance metric. To be specific, in a HetNet with $K$ base terminal stations (BTS's), we determine the optimal partition of the spectrum into $2^K$ possible spectrum sharing combinations. We use an interactive queueing model to characterize the flow level performance, where the service rates are decided by the spectrum partition. The spectrum allocation problem is formulated using a conservative approximation, which makes the optimization problem convex. We prove that in the optimal solution the spectrum is divided into at most $K$ pieces. A numerical algorithm is provided to solve the spectrum allocation problem on a slow timescale with aggregate traffic and service information. Simulation results show that the proposed solution achieves significant gains compared to both orthogonal and full spectrum reuse allocations with moderate to heavy traffic.
\end{abstract}

\section{Introduction}
\label{sec:Intro}
Densely deploying small cells is one of the major techniques to address the scarcity of spectrum resources for future cellular networks~\cite{CavAgr05MWC}. By reducing the coverage of each micro/pico base terminal stations (BTS), the capacity of a network can be significantly increased. With overlapping cells of all sizes, a heterogenous network (HetNet) almost always operates in the interference limited regime. Small cells may also lead to more pronounced traffic variations. Hence traditional frequency reuse and cell planning can be inefficient. Resource allocation according to traffic and demand conditions becomes highly desirable.

One conventional spectrum allocation scheme is fractional frequency reuse (FFR). Dynamic FFR can improve the sum-rate and total throughput of a network through effectively mitigating inter-cell interference~\cite{StoVis08INFOCOM,ChaTao09ICC,AliLeu09TransWC}. However these results are based on the assumption that traffic is backlogged at all BTS's. In a dense HetNet with dynamic traffic, physical layer performance metrics are not good indicators of user QoS, especially the delay experienced by a user. In this work, we focus on improving the average packet delay. With dynamic traffic, each BTS alternates between service and vacation periods, which introduces complicated inter-cell interference. A recent study in~\cite{DhiGan2012arXiv} points out that the backlogged traffic assumption, i.e., `always transmitting/interfering', will exaggerate the contribution of each BTS to inter-cell interference. Hence we propose a traffic aware interference model to address this issue. The optimization problems in~\cite{StoVis08INFOCOM,ChaTao09ICC,AliLeu09TransWC} are considered on fast timescales with instantaneous information exchange. Since traffic variation happens on a slower timescale in practice (compared to the timescale of scheduling) and there are limitations on the frequency with which a central controller can acquire traffic and service information from all BTS's, the spectrum allocation problem in this paper is solved by a central controller on a slow timescale as in~\cite{MadBor10JSAC,ZhuGuo2012Allerton} with aggregate traffic and service information.

In this work, we study resource allocation in dense deployed HetNets with traffic variations. The objective is to improve user QoS through spectrum resource adaptation on a slow timescale, as the long-term spatial load changes. The key is to link the spectrum and power resources in the physical layer to the QoS in the network layer. To do this, we use the service rates of the queues at all BTS's as this link. To be specific, a given spectrum allocation across cells induces an interference pattern and corresponding spectral efficiency on each segment of the spectrum. The spectral efficiencies along with the widths of the different segments of spectrum determine the service rate of each BTS, according to Shannon's capacity formula. The average packet delay at a given BTS is then determined by the service rate and the packet arrival rate. Thus an optimization problem can be formulated with the QoS as the objective, the spectrum allocation as desired variable, and the service rates as intermediate variables. Hence we consider the resource allocation problem as a joint physical layer and network layer optimization problem integrating results from both information theory and queueing theory.

The queueing scenario in our proposed system can be modeled by interactive queues. In general, it is difficult to get a closed-form steady state distribution of interactive queues. We use a conservative upper bound as an approximation of the average packet sojourn time. This upper bound is obtained by serving each queue independently at a conservative rate, which can be sustained regardless of the state of the other queues~\cite{BonBor2004ACM}. The spectrum allocation problem is formulated as a convex optimization problem, using the conservative upper bound. The geometric nature of the problem induces a special structure of the optimal solution. An efficient algorithm is provided to solve the optimization problem. Because of the use of conservative upper bound as objective, the optimum can be regarded as a QoS guarantee. According to the numerical results, the optimal solution greatly reduces the average packet delay in the heavy traffic regime compared to both orthogonal and full spectrum reuse allocations. The performance gain is observed mainly because each BTS is driven to allocate the right amount of spectrum to serve its own traffic demands, while leaving enough lightly loaded spectrum for adjacent BTS's.


\section{System Model}
\label{sec:SysMod}
We consider downlink data transmission in a HetNet with $K$ randomly deployed BTS's. Denote the set of all BTS's as $S=\{1,\dots,K\}$. The $K$ BTS's collectively share a unit bandwidth to serve their respective user equipments (UEs). A central controller determines which part of the spectrum is allocated to each BTS. Assuming the frequency resources are interchangeable, the problem is equivalent to deciding the bandwidth shared by each subset of BTS's, denoted by a $2^K$-tuple: $\boldsymbol{x}=\left(x(B)\right)_{B\in 2^S}$, where $2^S=\{B | B\subseteq S\}$ is the power set of $S$ containing all combinations of BTS's (including the empty one), and $x(B)\in[0,1]$ is the fraction of spectrum shared by BTS's in set $B$. Clearly, $\sum_{B\in2^S}x(B)=1$, and any efficient allocation would set $x(\emptyset)=0$. For example, if $K=2$, then there are basically three variables to decide on, $x(\{1\})$, $x(\{2\})$ and $x(\{1,2\})$, which denote the amount of spectrum allocated to BTS 1, BTS 2 exclusively and that shared by the two BTS's.

We next specify the relationship between spectrum allocation and the service rate at each BTS. The actual service rate in each cell depends on its own spectrum usage as well as the interference from other cells. To characterize the interaction among multiple BTS's, we define $A(t)$ as the set of BTS's that are transmitting data to their UEs at time $t$. Since we are interested in the average performance over a slow timescale, the time index $t$ will be omitted. The spectral efficiency of BTS $i$ on a segment of spectrum denoted by $s_i(C)$ is a function of the set $C$, which contains the BTS's actively sharing it. The service rate in cell $i$ when the active set is $\mathcal{A}$ is given by:
\begin{equation}
\label{eq:ServiceRate}
r_i(A)=\sum_{B\in 2^S}s_i(B\cap A)x(B).
\end{equation}
The intersection of sets in~\eqref{eq:ServiceRate} is because among all BTS's in $B$, only those in $B\cap A$ are transmitting. For concreteness, let the spectral efficiency of BTS $i$ with active set $C$ be calculated as:
\begin{equation}
\label{eq:SpeEff}
s_i(C)=\mathbb{1}(i\in C)\log\left(1+\frac{p_i l_{i}}{I_{i}(C)+n_i}\right),
\end{equation}
where $\mathbb{1}$ is the indicator funciton, $p_i$ is the constant transmit power spectral density (PSD), $l_i$ is the signal attenuation from BTS $i$ to its UEs, which includes both pathloss and slow-time fading, $I_{i}(C)$ is the constant interference PSD when BTS's in $C$ are generating interference, and $n_i$ is the noise PSD. The actual values of $s_i(C)$'s depend on transmit PSD at each BTS, the path-loss model and network topology. The receivers of a cell are assumed to be at the same point to simplify the interference model. The model, however, can be refined to address the locations of UEs by considering finer classification of UEs within each cell. Since the optimization will be performed on a slow timescale, this information, i.e., $s_i(C)$'s are assumed known by the central controller \textit{a priori}.

User service requests are modeled as packet arrivals at each BTS following a Poisson process with rate $\lambda_i$ at BTS $i$. All packet lengths are \textit{i.i.d.}\ exponentially distributed with unit mean. The objective is to minimize the average mean packet sojourn time through spectrum allocation, i.e., optimizing $\{x(B),~B\in 2^S\}$ for given $\lambda_i,~s_i(C),~\forall i\in S,~\forall C\in 2^S$. To evaluate the flow-level performance, we assume user requests within a cell are processed according to a `first come first serve' criterion.

The $K$-BTS network described above is a network of $K$ interactive queues, where the service rate of each queue depends on the status of the other queues at the same time. Such an interactive queueing system is also referred to as a \emph{coupled-processors} model. In the special case of two coupled queues, finding the steady-state joint distribution can be formulated as a Riemann-Hilbert problem~\cite{FayIas1979Springer}. Two coupled processors with generally distributed service times have been studied in~\cite{CohBox2000Elsevier}, which shows the joint workload distribution is the solution to a boundary value problem. These results are difficult to use for numerical computation. Also, few results exist for more than two coupled queues. A numerical method for solving the average packet delay using semidefinite program has been proposed in~\cite{RenCar2008}, which is again difficult to incorporate in our optimization. Here we use an upper bound on the true delay as the objective in the proposed optimization problem. The upper bound is achieved by decoupling the interactions among BTS's, which can be written in a simple closed form. Although the approximation is pessimistic, the effect on packet delay due to load variation is preserved.

\section{The Spectrum Allocation Problem}
\label{sec:ProbForm}
In this section, we introduce the upper bound of the mean sojourn time. The key is to assume each BTS always transmits at the worst-case rate, which is achievable regardless of other BTSs' states. This assumption is equivalent to assuming other BTS's are always backlogged and causing interference. Hence each BTS $i$ serves its users with constant rate $r_i(S)$, which is given by~\eqref{eq:ServiceRate} with $A=S$. From now on, we will use $r_i$ to denote $r_i(S)$ for simplicity. Therefore the $K$ interactive queues become $K$ independent $M/M/1$ queues with arrival rate $\lambda_i$ and service rate $r_i$ at BTS $i$. The mean packet sojourn time at BTS $i$ takes a simple form~\cite{Nel95Spinger}:
\begin{equation}
\label{eq:IndDelay}
t_i=\frac{1}{r_i-\lambda_i}.
\end{equation}

\subsection{Optimization Problem}
\label{subsec:Opt}
The spectrum allocation problem using the conservative approximation~\eqref{eq:IndDelay} is formulated as:
\begin{subequations}
\label{eq:SAP-Ind}
\begin{align}
\underset{\{x(B),~B\in 2^S\}}{
\text{minimize}}~&~ \frac{\lambda_i}{\sum_{j=1}^{k}\lambda_j}\sum_{i=1}^k \frac{1}{r_i-\lambda_i} \label{eq:Obj-Ind}\\
\text{subject to}~~&~r_i=\sum_{B\in 2^S}s_i(B)x(B),~\forall i\in S\label{eq:minServiceRate}\\
&~r_i>\lambda_i,~\forall i\in S\label{eq:Con2-Ind}\\
&~ x(B)\geq0,~\forall B\in 2^S\label{eq:Con1-Ind}\\
&~ \sum_{B\in 2^S}x(B)=1\label{eq:Con3-Ind}.
\end{align}
\end{subequations}
The objective~\eqref{eq:Obj-Ind} is the weighted average mean packet sojourn time of the entire network, where the weight $\frac{\lambda_i}{\sum_{j=1}^{k}\lambda_j}$ is the fraction of total traffic served by BTS $i$. Constraints in~\eqref{eq:Con2-Ind} are to guarantee the stability of the queues. The optimization problem~\eqref{eq:SAP-Ind} is a convex optimization problem. To see this, we can take $x(B)$'s and $r_i$'s as the optimization variables. Thus all the constrains are linear, and the objective is a linear combination of convex functions as $t(r)=1/(r-\lambda)$ is convex in $r$ on $(\lambda,\infty)$.

The convex optimization problem~\eqref{eq:SAP-Ind} has a unique globally optimal solution~\cite{BerDim1999nonlinear}. Due to the special geometric structure of~\eqref{eq:SAP-Ind}, the optimal solution has the following property.
\begin{theorem}
\label{thm:KBTS}
The optimal solution of the $K$-BTS spectrum allocation problem divides the spectrum into at most $K$ segments:
\begin{equation}
\nonumber
\left|\{B~|~x(B)>0,~B\in2^S\}\right|\leq K.
\end{equation}
\end{theorem}

\begin{IEEEproof}
\label{pf:KBTS}
Denote the rate vector $\rr$ and spectral efficiency vector of sharing combination $B$, $\ssv(\mathcalB)$, as $\rr=[r_1,\dots,r_K]$ and $\ssv(\mathcalB)=[s_1(\mathcalB), \dots, s_K(\mathcalB)]$. According to~\eqref{eq:minServiceRate} to \eqref{eq:Con3-Ind}, $\rr\in\mathds{R}_+^K$ is a convex combination of the $2^K-1$ points $\{\ssv(\mathcalB)~|~\mathcalB\in2^\mathcalK\}$, with coefficients $\{x(\mathcalB)~|~\mathcalB\in2^\mathcalK\}$, i.e., $\rr=\sum_{\mathcalB\in2^\mathcalK}\ssv(\mathcalB)x(\mathcalB)$. In other words, any $\rr$ given by~\eqref{eq:minServiceRate} is in the convex hull of $\{\ssv(\mathcalB)~|~\mathcalB\in2^\mathcalK\}$. By Carath\'eodory's Theorem~\cite{Egg69Convexity}, $\rr$ lies in a $d$-simplex with vertices in $\{\ssv(\mathcalB)~|~\mathcalB\in2^\mathcalK\}$ and $d\leq K$, i.e., $\rr$ can be written as a convex combination of $x(\mathcalB)$'s with at most $K+1$ nonzero $x(\mathcalB)$'s. This holds for any $\rr$ satisfying~\eqref{eq:minServiceRate} to \eqref{eq:Con3-Ind}. Furthermore, the $\rr^*$ corresponding to the optimal solution to~\eqref{eq:SAP-Ind} must be Pareto optimal in terms of the rate allocation, i.e., there is no spectrum allocation $\boldsymbol{x}$ such that $r_i^*\leq \sum_{\mathcalB\subseteq\mathcalK} s(\mathcalB)x(\mathcalB),~\forall i\in\mathcalK$ and at least one inequality is strict. This is because any spectrum allocation that could increase the rate at any BTS without decreasing the rates at other BTS's would also decrease the objective~\eqref{eq:Obj-Ind}. Hence $\rr^*$ cannot be an interior point of the $d$-simplex, and must lie on some $m$-face of the $d$-simplex with $m<d\leq K$. Therefore $\rr^*$ can be written as a convex combination of $m+1\leq K$ nonzero $x(\mathcalB)$'s.
\end{IEEEproof}

The geometric interpretation can be generalized to any subset of the $K$ BTS's.
\begin{corollary}\label{cor:Sub}
  The optimal solution of the $K$-BTS spectrum allocation problem divides the spectrum exclusively used by any subset $\mathcalM\subseteq\mathcalK$ of the $K$ BTS's with $m$ BTS's into at most $m$ segments:
\begin{equation}
\nonumber
\label{eq:Cor}
\left|\{\mathcalB~|~x(\mathcalB)>0,~\mathcalB\in2^\mathcalM\}\right|\leq m.
\end{equation}
\end{corollary}

\begin{IEEEproof}
\label{pf:sub}
If the bandwidths of the spectrum segments not in $2^\mathcalM$ are fixed at their optimal values, then~\eqref{eq:SAP-Ind} becomes an optimization problem of variables $x(\mathcalB),~\mathcalB\in2^\mathcalM$, the service rates at BTS's not in $\mathcalM$ are fixed and the service rates for the $m$ BTS's in $\mathcalM$ are convex combinations of $x(\mathcalB),~\mathcalB\in2^\mathcalM$ plus a constant vector in $\mathds{R}_+^m $. Therefore Corollary~\ref{cor:Sub} can be proved using a similar argument as in the proof of Theorem~\ref{thm:KBTS}.
\end{IEEEproof}
%

\subsection{An Efficient Algorithm}
\label{subsec:EA}
The structure of the optimal solution given by Theorem~\ref{thm:KBTS} suggests the possibility of solving~\eqref{eq:SAP-Ind} more efficiently in real systems. Originally we needed to determine the sizes of $2^K-1$ spectrum segments. The computational complexity is exponential in $K$ using a standard convex optimization algorithm. By Theorem~\ref{thm:KBTS}, we only need to decide the sizes of the $K$ nonzero segments. The difficulty is that we do not know which $K$ pieces out of the $2^K-1$ possibilities. Algorithm~\ref{alg:SAP-Ind} is an iterative algorithm to efficiently solve the $K$-BTS spectrum allocation problem.

\begin{algorithm}
\caption{Iterative algorithm for solving problem~\eqref{eq:SAP-Ind}}
\label{alg:SAP-Ind}
\begin{algorithmic}[]
\label{alg:SAP-Ind}
\INPUT {$\lambda_i$ and $s_i(C)$ for all $i\in \mathcalK$ and $C\in2^\mathcalK$.}
\OUTPUT{$x(\mathcalB)$ for all $\mathcalB\in2^\mathcalK$}.
\Init{Find a feasible solution $x_0(\mathcalB)$ by solving~\eqref{eq:SAP-Ind} with constant objective.
$\Nc = \{\mathcalB~|~x_0(\mathcalB)>0\}$, $\Nc'=\emptyset$, $\Nc^+ = 2^\mathcalK$.}
\While{$\Nc^+\not\subseteq \Nc'$}
    \State 1. $\Nc'=\Nc$;
    \State 2. Find $x(\mathcalB)$ by solving~\eqref{eq:SAP-Ind} starting from $x_0(\mathcalB)$ with the additional constraints, $x(\mathcalB)=0,~\forall\mathcalB\notin \Nc$;
    \State 3. Compute the partial derivatives of the objective function~\eqref{eq:Obj-Ind} with respect to all $x(\mathcalB)$'s, $\Delta_{x(\mathcalB)}=-\sum_{i\in\mathcalB}\frac{\lambda_is_i(\mathcalB)}{(r_i-\lambda_i)^2}$;
    \State 4. $\Nc^+=\{\mathcalB~\text{for}~K~\text{smallest}~\Delta_{x(\mathcalB)}\}$, $\Nc=\Nc\cup \Nc^+$, $x_0(\mathcalB)=x(\mathcalB)$.
\EndWhile
\end{algorithmic}
\end{algorithm}

The algorithm requires to start at a feasible point. A standard method is to solve a modified version of optimization problem~\eqref{eq:SAP-Ind} by replacing the objective function~\eqref{eq:Obj-Ind} with a constant. The resulting optimization problem can be transformed to a linear program in standard form, which can be solved using the simplex method~\cite{BerDim97LP}.

Starting from a feasible point $\left(x_0(\mathcalB)\right)_{\mathcalB\in2^\mathcalK}$, let $\Nc$ be the candidate set, which initially includes the indices of those nonzero spectrum segments of the initial point. In each iteration, the algorithm finds the optimal solution within the candidate set $\Nc$. After each iteration, the partial derivatives with respect to all $x(\mathcalB)$'s are calculated (including those not in $\Nc$). The $K$ segments with the $K$ smallest derivatives are added to the candidate set. (The number of variables added to the candidate set may be less than $K$, since some of the $\mathcalB$'s of the $K$ smallest derivatives may already be in $\Nc$.) The algorithm continues with the solution found in the last iteration as the new initial point and the expanded candidate set, until the candidate set stops growing. At the end of each iteration, if the solution is not optimal, there must be some $\mathcalB$'s outside the candidate set that have smaller partial derivatives. Since we only add more $\mathcalB$'s to the candidate set with each iteration, in the worst-case, the candidate set will eventually include all $2^K-1$ variables. Hence the proposed algorithm is guaranteed to converge to the global optimum.

\begin{figure}
\centering
\includegraphics[width=2.8in]{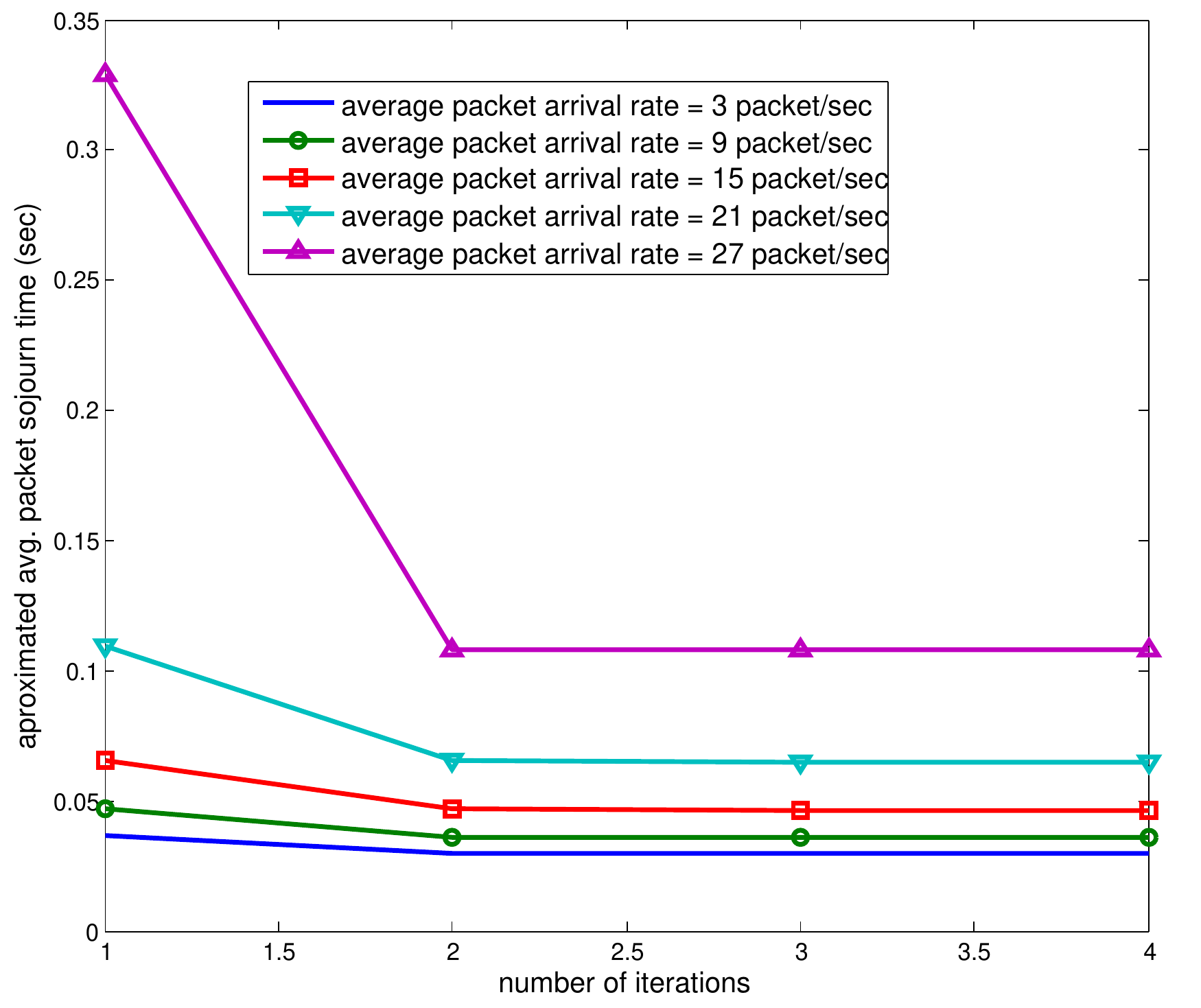}
\caption{Approximated average 
delay versus number of iterations using Algorithm~\ref{alg:SAP-Ind} with different average packet arrival rates.}
\label{fig:Alg}
\end{figure}

The algorithm is more efficient when starting at an initial point with fewer nonzero spectrum segments. Therefore we can always use the full-spectrum-reuse allocation, $x(\mathcalK)=1$, as an initial point if it is feasible. Even if it is not, the solution obtained by the initialization method has no more than $K+1$ nonzero spectrum segments according to the properties of basic feasible solution to a linear program~\cite{BerDim97LP}. Examples of delay versus number of iterations are shown in Fig.~\ref{fig:Alg} with $K=7$ and different traffic loads.
In the simulation, Algorithm 1 starts with the full-spectrum-reuse allocation. The figure shows that the algorithm converges within a few iterations.

\section{Numerical Results}
\label{sec:Sim}
\subsection{Performance of the Spectrum Allocation Method}
\label{subsec:Sim}
In the simulation, we adopt the quantized HetNet model in~\cite{ZhuGuo2012Allerton}. A $100\times100~\text{m}^2$ area is quantized by hexagons with distance between the centers of adjacent hexagons being $20~\text{m}$. Seven BTS's are uniformly randomly dropped at the vertices of the hexagons. UE locations within each hexagon are approximated by the center of the hexagon. UEs are assigned to their respectively nearest BTS's. If there is a tie, UEs in the hexagon are equally distributed to the nearest BTS's. We only consider path-loss in this simulation, although slow fading can be easily incorporated. The average spectral efficiency of BTS $i$ in~\eqref{eq:SpeEff} is calculated as the mean of the spectral efficiencies in the hexagons served by BTS $i$. Other parameters used in the simulation include: path-loss exponent is 3; transmit PSD is 1 watt/Hz for all BTS's; noise PSD is 0.125 micro-watt/Hz.

\begin{figure}
\centering
\includegraphics[width=2.8in]{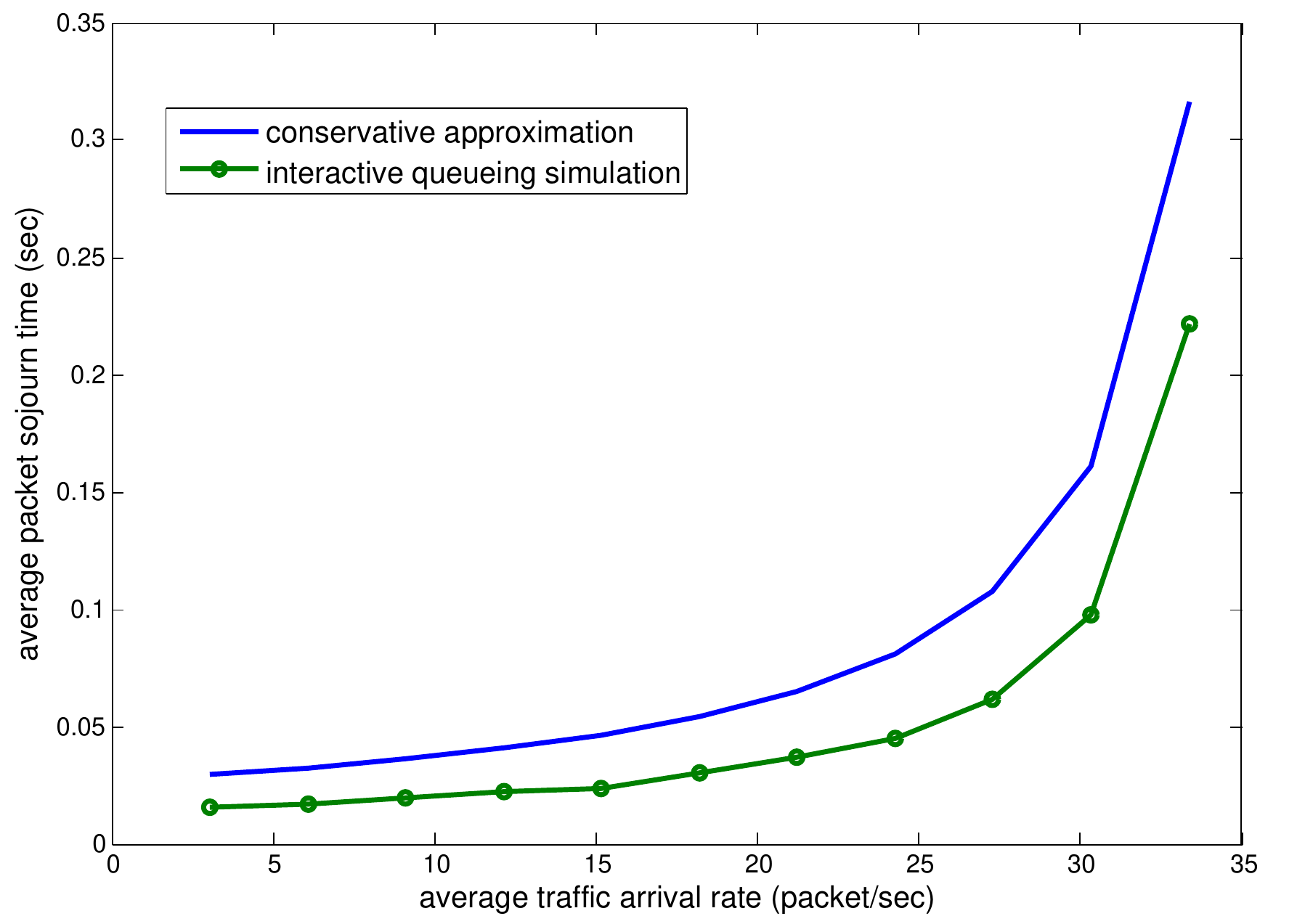}
\caption{Comparison of average packet sojourn time versus average packet arrival rate using the approximation and simulation.}
\label{fig:AprxSimu}
\end{figure}
The approximate mean sojourn time in~\eqref{eq:Obj-Ind} and the real mean sojourn time obtained by simulating the interactive queues using the uniformization method are shown in Fig.~\ref{fig:AprxSimu} for different traffic loads. Both the approximated delay and the real delay have similar trends as traffic increases.

\begin{figure}
\centering
\includegraphics[width=2.8in]{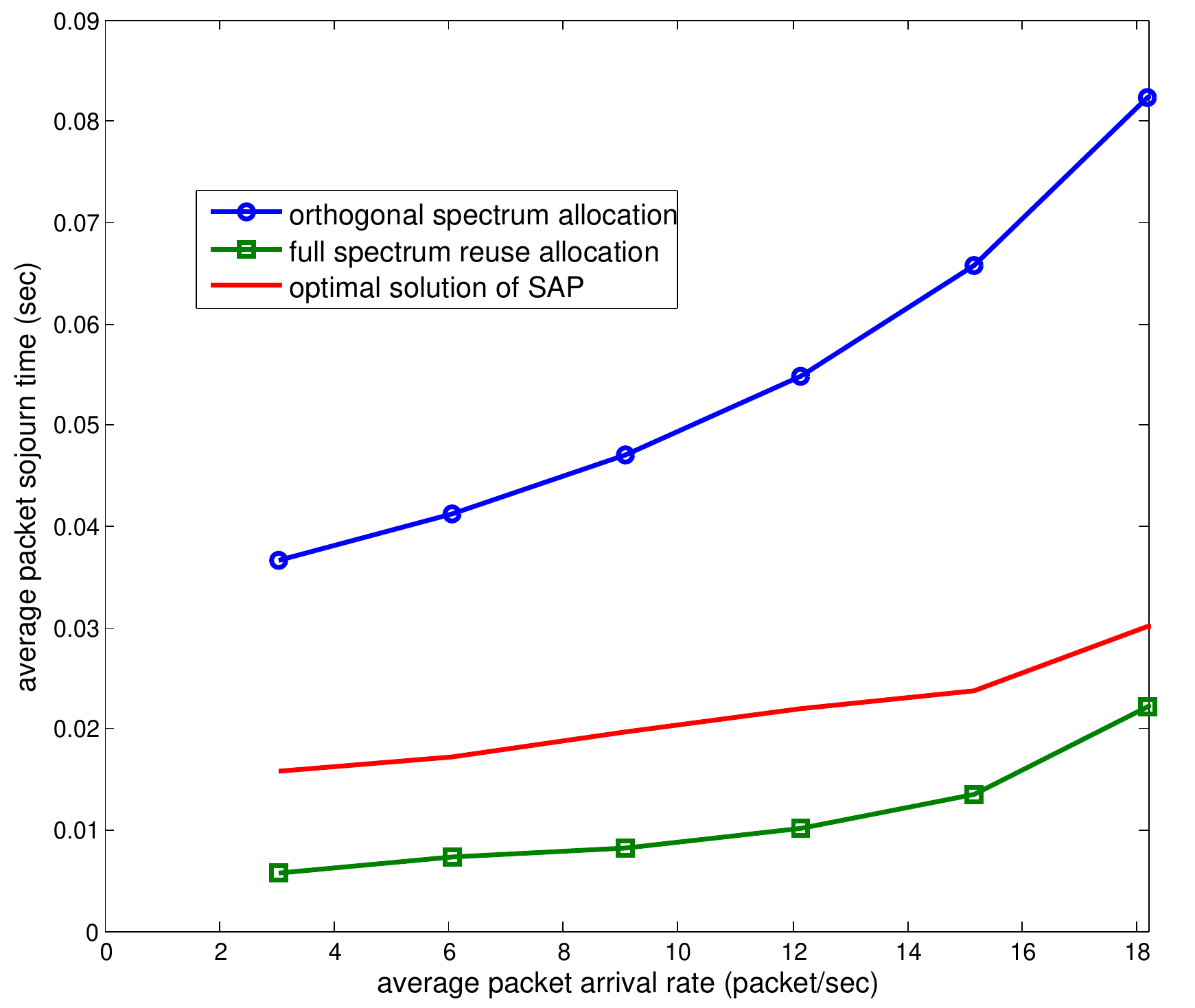}
\caption{Simulated average packet sojourn time using orthogonal spectrum allocation, full spectrum reuse allocation and the optimal solution of the spectrum allocation problem in light traffic.}
\label{fig:Light}
\end{figure}
\begin{figure}
\centering
\includegraphics[width=2.8in]{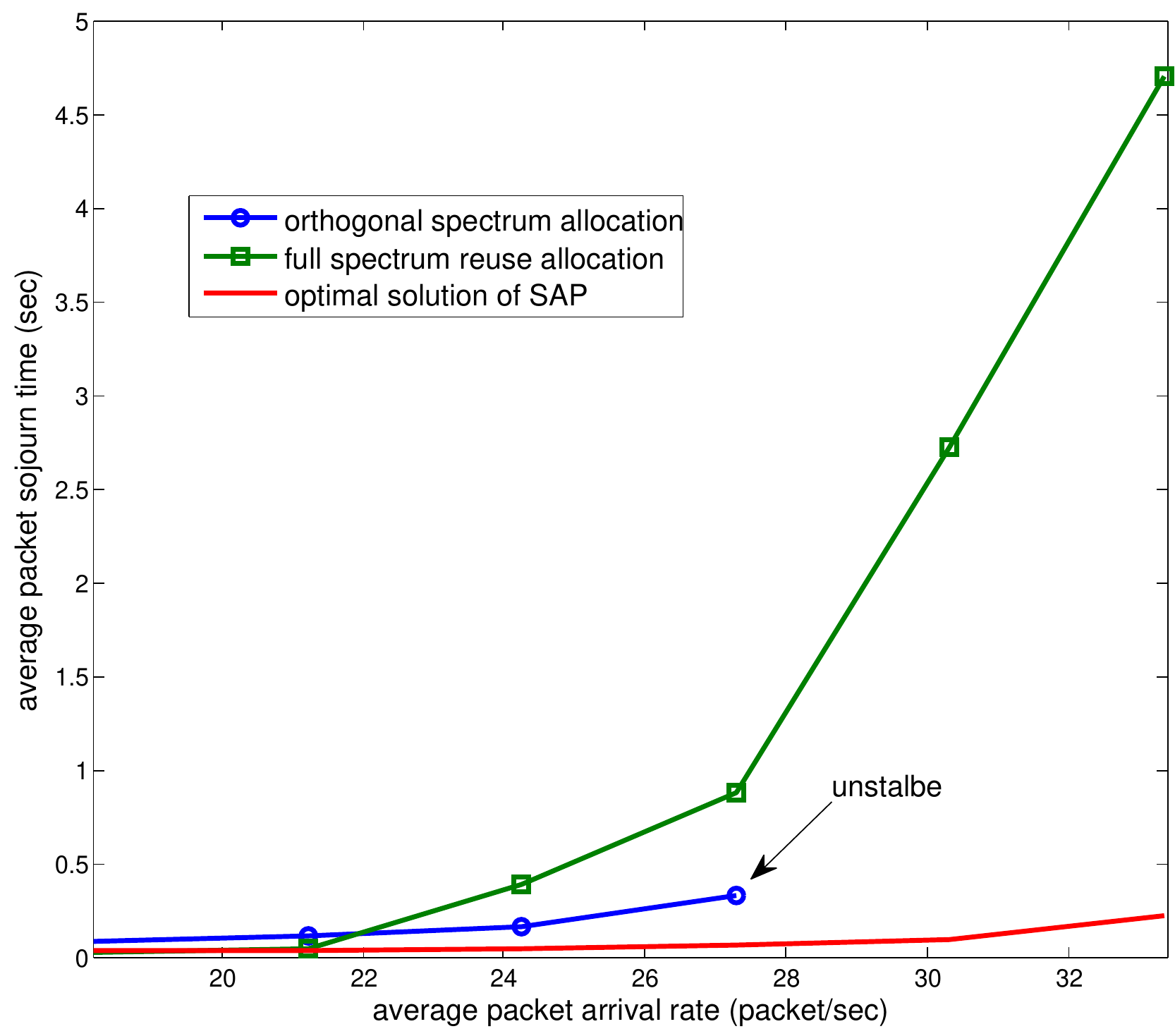}
\caption{Simulated average packet sojourn time using orthogonal spectrum allocation, full spectrum reuse allocation and the optimal solution of the spectrum allocation problem in heavy traffic.}
\label{fig:Heavy}
\end{figure}
The optimal orthogonal spectrum allocation and the full spectrum reuse allocation are compared with the optimal solution in Fig.~\ref{fig:Light} and Fig.~\ref{fig:Heavy}.
Fig.~\ref{fig:Light} shows the comparison in the light traffic regime. The average packet delay given by the optimal solution is between those given by the optimal orthogonal spectrum allocation and the full spectrum reuse allocation. Since traffic load is light, BTS's will have a larger fraction of time being empty. Hence the worst case transmit rate assumption exaggerates the harm of inter-cell interference. Due to rare concurrent transmissions among the BTS's, the full spectrum reuse allocation turns out to be more efficient. However, as the average packet arrival rate increases, the optimal solution of~\eqref{eq:SAP-Ind} outperforms the other two spectrum allocation schemes as shown in Fig.~\ref{fig:Heavy}. Due to increasing impact of inter-cell interference, the optimal orthogonal spectrum allocation becomes more efficient than the full spectrum reuse allocation as well. However, after the average packet arrival rate reaches 27 packets/sec, there is no orthogonal spectrum allocation that maintains the stability of the system. On the other hand, the optimal solution of~\eqref{eq:SAP-Ind} remains stable. Since the optimization problem exhaust all spectrum sharing possibilities, traffic and topology driven spectrum reuse is realized in the heavy traffic regime.

\begin{figure}
\centering
\includegraphics[width=2.8in]{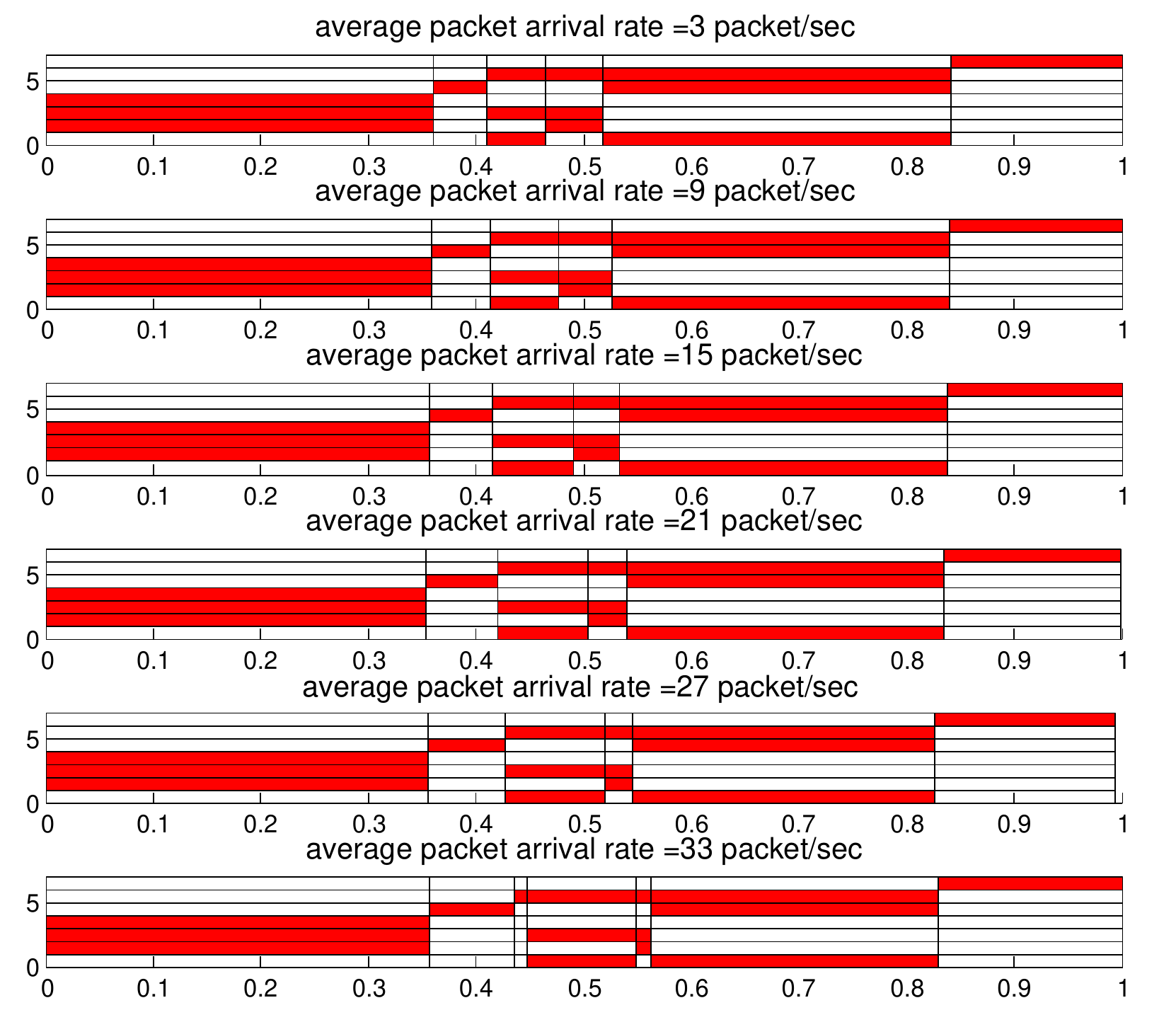}
\caption{Optimal solution of the spectrum allocation problem for different average packet arrival rates.}
\label{fig:OptSol}
\end{figure}
The optimal solutions for different average packet arrival rates are shown in Fig.~\ref{fig:OptSol}. Each subplot in the figure is the spectrum partition for a specific average packet arrival rate, with only the nonzero segments. Each row of a subplot is the spectrum usage of the corresponding BTS. Each BTS only transmits on the shaded pieces of spectrum in its row. By counting the number of pieces in the partition, we can verify Theorem~\ref{thm:KBTS}. Note that in the light traffic regime, the optimal solution still orthogonalizes the spectrum usage among BTS's to some extent, which is why the optimal solution is worse than the full spectrum reuse allocation in this regime. (This is due to the conservative approximation of the true objective.)


\subsection{Power Control}
\label{subsec:Power}
\begin{figure}
\centering
\includegraphics[width=2.8in]{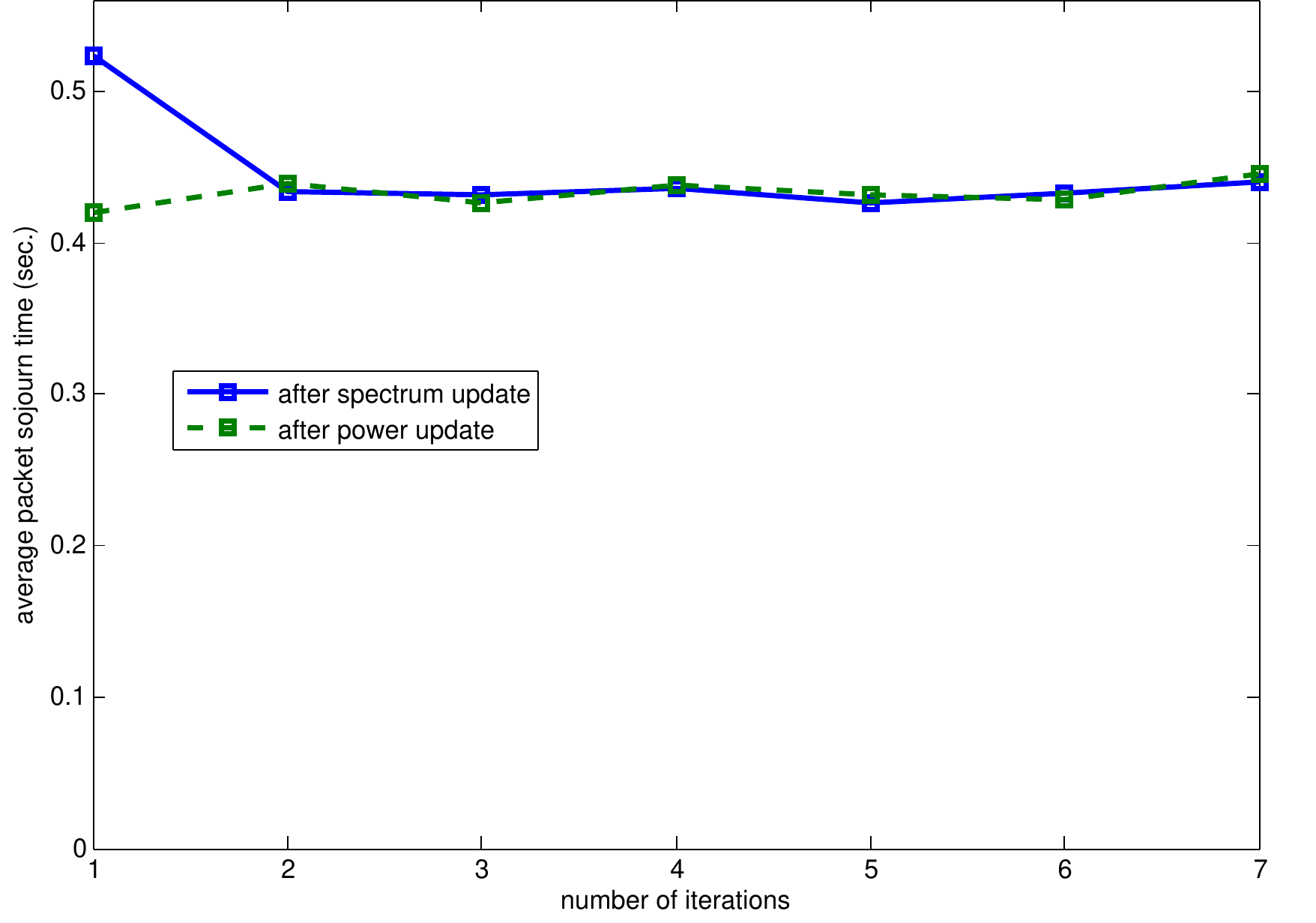}
\caption{Average packet sojourn time versus number of iterations.}
\label{fig:DelayIteration}
\end{figure}

All the discussions and results so far are based on the assumption that the spectral efficiencies under different partitions are fixed. This assumption helps us simplify the relationship between the spectrum allocation and the service rates. In fact, the service rates are linear functions of the spectrum variables. The assumption is true if all BTS's transmit with fixed power spectral density. However, in practice, we may have a fixed total transmit power constraint at each BTS instead. Hence power control becomes an issue. For example, the spectral efficiencies should be higher for the orthogonal allocation, since each BTS could concentrate all its power on the piece of spectrum exclusively used by itself.

The joint power control and spectrum allocation problem is in general more complicated. Hence we provide a simple solution using alternative spectrum and power updates. We start with fixed $s_i(C)$ assuming each BTS uniformly allocates its maximum transmit power across the whole spectrum. We iterate between the following two updates:
\begin{itemize}
\item[1.] Update the spectrum allocation $x(B),~\forall \mathcal{B}\in2^S$ with the current $s_i(C),~\forall i\in S,~\forall C\in2^S$ by solving the spectrum allocation problem.

\item[2.] Update the spectral efficiencies $s_i(\mathcal{C}),~\forall i\in S,~\forall C\in 2^S$  with the current $x(B),~\forall B\in 2^S$ by letting each BTS $i$ uniformly allocate its maximum transmit power over the spectrum it uses, which includes $x(B)$'s, with $i\in B$.
\end{itemize}
The iteration terminates unti $x(B),~\forall B\in 2^S$ converges (This is not guaranteed). The average packet sojourn time after the spectrum allocation update and the spectral efficiency update at each iteration is shown in Fig.~\ref{fig:DelayIteration} for an average packet arrival rate per BTS of 24 packets/second. The figure shows that the delay performance converges very quickly. The mean sojourn times decrease substantially after the first spectral efficiency update. This is because at this average packet arrival rate both allocations orthogonalize the spectrum use among neighboring BTS's to some extent. This kind of convergence behavior can be expected in general, since spatial reuse will occur in the optimal solution. With each BTS using a fairly large amount of the spectrum, the spectral efficiencies will not change much after several iterations.


Although all BTS's have the same transmit PSD in the simulations, the proposed spectrum allocation problem and algorithm can be directly applied to HetNets with arbitrary power, topology and traffic conditions.

\section{Conclusion}
\label{sec:Con}
We have formulated a joint physical layer and network layer optimization problem to minimize average delay in HetNets using the combination of information theory and queueing theory. The optimization problem takes traffic arrival rates and spectral efficiencies as the input and gives the spectrum partition as the output. Numerical results obtained by simulating the interactive queues suggest the optimal solution of the proposed spectrum allocation achieves a significant delay performance gain in the heavy traffic regime, compared to both orthogonal and full spectrum reuse allocations. For future work, we plan to look for a more accurate approximation of the average packet sojourn time in order to improve the performance in the light and moderate traffic regimes, since the conservative approximation presented has ignored the impact of traffic variations on inter-cell interference.

\bibliographystyle{IEEEtran}
\bibliography{def,IEEEabrv,ref}
%
%
%

\end{document}